\documentclass{PoS}
\usepackage{amsmath}

\newcommand{\Tr}{\operatorname{Tr}}

\title{Emergent Geometries from the BMN Matrix Model}

\ShortTitle{Emergent Geometries from the BMN Matrix Model}

\author{\speaker{Yuhma Asano}\thanks{This paper is based on Ref.~\cite{Asano:2018nol}, written with Veselin Filev, Samuel Kov\'a\v{c}ik and Denjoe O'Connor, and on collaborations with Goro Ishiki, Takashi Okada, Shinji Shimasaki and Seiji Terashima.}\\
        KEK Theory Center, High Energy Accelerator Research Organization, \\
	1-1 Oho, Tsukuba, Ibaraki 305-0801, Japan\\
        E-mail: \email{yuhma@post.kek.jp}}
\abstract{
We review recent results of emergent geometries in the BMN matrix model,
a one-dimensional gauge theory considered as a non-perturbative formulation 
of M-theory on the plane-wave geometry.
A key to understand the emergent geometries is 
the eigenvalue distribution of a BPS operator.
Gauge-theory calculation shows that the BPS operator reproduces 
the corresponding supergravity solutions in the gauge/gravity duality
and also brane geometries in the M-brane picture.
At finite temperatures, these geometries should be realised in a non-trivial way.
Monte Carlo simulations of this gauge theory 
revealed two types of phase transitions:
the confinement/deconfinement transition and the Myers transition,
which provide insights into the emergence of the geometries.
Especially, the numerical results qualitatively agree with 
the critical temperature of the confinement/deconfinement transition
predicted on the gravity side.
\\
\\
Preprint number: KEK-TH-2211
}

\FullConference{Corfu Summer Institute 2019 "School and Workshops on Elementary Particle Physics and Gravity" (CORFU2019)\\
		31 August - 25 September 2019\\
		Corfù, Greece}

\begin{document}

\section{Introduction}
Matrix models are hopeful candidates of non-perturbative formulation of string theory.
They are obtained by the matrix regularisation of 
superstring worldsheet or supermembrane worldvolume theory,
and the resultant theories are lower dimensional gauge theories,
such as the BFSS matrix model \cite{deWit:1988wri,Banks:1996vh}, 
which is one-dimensional
and the IKKT matrix model \cite{Ishibashi:1996xs}, 
which is zero-dimensional.
This %fact 
suggests higher dimensional geometries appearing in superstring theory
should emerge from the lower dimensional theory.
In fact, classical configurations in the matrix models are
considered as D-brane configurations,
some of which have higher spatial dimensions.
Another example is that
the gauge/gravity duality conjecture relates
the BFSS matrix model to a black 0-brane solution 
in the ten-dimensional supergravitational theory.

%HERE
% The emergence of geometries

The BMN matrix model \cite{Berenstein:2002jq}, 
also known as the plane-wave matrix model,
% especially fits the idea of
is especially suitable for the purpose of understanding
the emergence of higher dimensional geometry.
This is a mass-deformed version of the BFSS matrix model
which keeps the original number of dynamical supercharges\footnote{
There are other 16 kinematical supercharges, which trivially shifts fermions,
to restore the entire superalgebra in type II string theories or M-theory.
}, 16;
and they form $SU(2|4)$ symmetry.
The action is in the form of
\begin{align}
 S&=N\int dt
 \Tr\Bigl[
 \frac{1}{2}(D_t X^a)^2
 +\frac{1}{2}(D_t X^m)^2
 -\frac{1}{4}\left( \frac{\mu}{3}\epsilon_{abc}X^c-i[X^a,X^b] \right)^2
 +\frac{1}{2}[X^a,X^n]^2
 \nonumber \\
 &\hspace{50pt}
 +\frac{1}{4}[X^m,X^n]^2
 -\frac{\mu^2}{72}X^mX^m
 +\text{fermions}
 \Bigr],
 \label{BMN-action}
\end{align}
% X^loc = 6/mu X^this
% t^loc = mu/6 t^this
% g^2N = (6/mu)^3
where % $a,b,c$ run through 1, 2, 3, and $m,n$ run from 4 to 9.
$X^a(t)$, $a=1,2,3$, and $X^m(t)$, $m=4,\cdots,9$, are $N\times N$ matrices,
and $D_t$ is the covariant derivative with an $SU(N)$ gauge field.
It explicitly shows $R\times SO(3)\times SO(6)$ symmetry\footnote{
$R$ is the translational symmetry in the direction of time.
% Note that the supersymmetric transformation is not invariant 
% under the $R$ symmetry.
},
which is the bosonic part of the universal cover of $SU(2|4)$ symmetry.
One of virtues of this matrix model is that
it is well controlled thanks to the mass parameter, $\mu$.
Without the mass-deformation, the BFSS matrix model has flat directions
in the potential
so that the moduli space of its vacua is infinite even at finite $N$.
In contrast,
the BMN matrix model has a finite number of discrete vacua at finite $N$.
They are labeled by representations of $SU(2)(\cong SO(3))$, %for $X^a$,
and all of them are exact quantum mechanical vacua \cite{Dasgupta:2002ru}.

Another remarkable feature of the BMN matrix model 
with regard to the emergent geometry
is the relation to 
four-dimensional $\mathcal{N}=4$ super Yang-Mills theory (SYM).
This matrix model can be obtained by the dimensional reduction of
$\mathcal{N}=4$ SYM on $\mathbb{R}\times S^3$
to one dimension.
Conversely,
the large-$N$ equivalence for the BMN matrix model is 
conjectured---the matrix model around a special vacuum 
reproduces $\mathcal{N}=4$ SYM 
in the large-$N$ limit \cite{Ishiki:2006yr,Ishii:2008ib}.
The equivalence was verified perturbatively in Ref.~\cite{Ishii:2008ib,Kitazawa:2008mx,Ishiki:2011ct}, 
non-perturbatively for a supersymmetric sector in Ref.~\cite{Asano:2012zt}
and numerically in Ref.~\cite{Ishiki:2008te,Ishiki:2009sg}.
% \cite{Honda:2011qk,Honda:2010nx}

The large-$N$ equivalence is consistent with 
a proposed gauge/gravity duality for the BMN matrix model and 
other $SU(2|4)$ symmetric gauge theories \cite{Lin:2005nh,Lin:2004nb}.
The corresponding gravitational theory is 
type IIA ten-dimensional supergravity, or equivalently, 
translationally invariant eleven-dimensional supergravity,
around half-BPS solutions.
The half-BPS solutions are characterised by 
droplets scattering on one-dimensional subspace in the gravitational theory,
where either quantised D2- or NS5-brane charge resides on each droplet,
and each of the solutions corresponds to 
a vacuum in $SU(2|4)$ symmetric gauge theories.
The consistency of the large-$N$ equivalence can be seen from
the relationship between the droplet configurations 
corresponding to $\mathcal{N}=4$ SYM on $\mathbb{R}\times S^3$
and to the BMN matrix model---the droplet configuration 
for $\mathcal{N}=4$ SYM on $\mathbb{R}\times S^3$
can be obtained from a certain limit of 
that for a special vacuum of the BMN matrix model.
Not only for $\mathcal{N}=4$ SYM on $\mathbb{R}\times S^3$,
can the droplet configurations for other $SU(2|4)$ symmetric gauge theories 
such as $\mathcal{N}=4$ SYM on $\mathbb{R}\times S^3/\mathbb{Z}_k$
and $\mathcal{N}=8$ SYM on $\mathbb{R}\times S^2$
be reproduced from the droplet configuration 
for vacua of the BMN matrix model,
and the corresponding large-$N$ equivalence has been checked \cite{Ishiki:2006yr,Ishii:2008ib,Kitazawa:2008mx,Ishiki:2011ct,Asano:2012zt}.
Thus, from this point of view,
the BMN matrix model can be considered as 
a ``master theory'' of the $SU(2|4)$ symmetric gauge theories, and 
each vacuum of the matrix model is dual to each of all the supergravity solutions.

Since supergravity solutions in this gauge/gravity duality 
correspond to supersymmetric vacua of the BMN matrix model,
% which are supersymmetric,
calculation in a BPS sector of the matrix model
should capture emergence of the dual supergravity solutions.
The application of the supersymmetric localisation technique \cite{Pestun:2007rz}
to the BMN matrix model \cite{Asano:2012zt}
revealed that the equations that determine the supergravity solutions 
are equivalent to the saddle point equations 
appearing in the matrix integral deduced by the localisation calculation
\cite{Asano:2014vba,Asano:2014eca}.
This equivalence will clarify how the supergravitational geometries
are embedded in the matrix model and the other gauge theories with $SU(2|4)$ symmetry.

The localisation calculation also revealed
the relationship to the brane picture in the gauge/gravity duality.
The duality is considered as 
a result of two ways to describe a bunch of branes---one way
is the field theory of strings on the branes and
the other is the classical gravitational force produced by the branes.
These pictures should be equivalent in the limit where
gravitation decouples from the brane system.
Thus the equivalence between the supergravity equations and 
the saddle point equations in the matrix model implies that
there should be a brane picture and 
that the saddle point equations would also describe 
the geometry of the branes.
It is seen from the localisation calculation that
some solutions on the matrix-model side realise the spherical geometry of
5-branes or 2-branes with zero light-cone energy in M-theory \cite{Asano:2017xiy,Asano:2017nxw}.

These two emergence phenomena of geometries\footnote{
Emergence of geometries related with the BMN matrix model
was also discussed in Ref.~\cite{Berenstein:2004kk,Takayama:2005yq,Berenstein:2005aa,Berenstein:2005jq}.
},
seen from the supersymmetric localisation calculation,
only tell how the geometries embed in the BMN matrix model.
On the other hand,
dynamics of the emergent geometries would 
require computation beyond the supersymmetric sector.
One promising way to achieve it is Monte Carlo simulation.
The rational hybrid Monte Carlo method is 
a standard approach to simulate such a supersymmetric matrix model.
In the context of the gauge/gravity duality,
numerical studies of supersymmetric gauge theories started in Ref.~\cite{Hanada:2007ti,Catterall:2007fp},
and evidence of the gauge/gravity duality for the BFSS matrix model
with $\alpha'$ and quantum ($1/N$) corrections
was obtained in
Ref.~\cite{Anagnostopoulos:2007fw,Catterall:2008yz,Hanada:2008ez,Hanada:2013rga,Kadoh:2015mka,Filev:2015hia,Berkowitz:2016jlq}
by measurement of the energy and related observables.
% $\alpha'$ and quantum ($1/N$) corrections and reproduction of coefficient in the leading term \cite{Hanada:2008ez,Hanada:2013rga,Berkowitz:2016jlq}
Moreover, the black 0-brane geometry on the gravity side of the BFSS model
was shown rather directly with a temporal Wilson loop \cite{Hanada:2008gy} and with correlation functions \cite{Hanada:2009ne,Hanada:2011fq}.

The rational hybrid Monte Carlo method was
also applied to the BMN matrix model.
The first full simulation was done in Ref.~\cite{Catterall:2010gf}; 
then results related with emergent geometries were provided
in Ref.~\cite{Honda:2013nfa,Asano:2018nol,Schaich:2020ubh}.
Especially Ref.~\cite{Asano:2018nol} showed
not only a result of the confinement/deconfinement transition qualitatively
consistent with the prediction from the gravity side \cite{Costa:2014wya}
but also a result directly related with transitions between different geometries.
The simulation showed, when the matrix size and the lattice size
are fixed to reasonable values,
a dominant vacuum is transitioned 
from the trivial vacuum, $X^a=0$, 
to a fuzzy-sphere vacuum with non-zero $SU(2)$ spins
as the temperature decreases.

This paper is organised as follows.
In section 2, we review the result of the localisation technique and 
see how the geometries in the supergravity picture and the brane picture
embed in the gauge theory.
Then in section 3,
we review the gravity side where the dual matrix model is thermal,
and show numerical results.
Finally we conclude and summarise them in section 4,
showing some relevant future aspects.

\section{Emergent Geometries}
In order to see the realisation of geometries,
let us consider a BPS operator,
\begin{equation}
 \phi(t)
  =X_3(t)
  +i\left(
     \sin(\tfrac{\mu}{6}t)X_8(t) + \cos(\tfrac{\mu}{6}t)X_9(t)
    \right).
\end{equation}
This is invariant under supersymmetry associated with four supercharges.
Insertions of this operator, $\phi$, at fixed $t$ break 
the bosonic symmetry $R\times SO(3)\times SO(6)$
down to $SO(2)\times SO(5)$.
Thus we naively guess $\phi$ would represent a direction 
perpendicular to $\mathbb{R}\times S^2\times S^5$, 
which corresponds to the bosonic symmetry,
and this should be the non-trivial direction in terms of the symmetry.

The supersymmetric localisation technique reduces
calculation for the vacuum expectation value of any functions of $\phi$
into much simpler matrix integration\footnote{
To be precise, the theory needs to be Wick-rotated first with $t=-i\tau$
for the localisation method.
} \cite{Asano:2012zt}:
\begin{equation}
 \left\langle
 \prod_I \Tr f_I(\phi(t_I))
 \right\rangle
 =\left\langle
 \prod_I \Tr f_I(-\tfrac{\mu}{3}L_3+\tfrac{i\mu}{6}M) %(\tfrac{\mu}{6}(-2L_3+iM))
 \right\rangle_{\!\!\! MM},
 \label{localisation_vev}
\end{equation}
where 
$L_a$ are the $SU(2)$ generators and $M$ is a ``moduli'' Hermitian matrix
satisfying $[L_a, L_b] = i\varepsilon_{abc}L_c$ and $[L_a,M] = 0$.
Here $\langle\cdots\rangle_{MM}$ is an expectation value computed by an integral of matrix $M$.
% Localising loci for this BPS sector is
% the configuration
% $X^a=-\frac{\mu}{3}L_a$, $X^9=\frac{\mu}{6\cos(\mu t/6)}M$ and $X^m=0$ for $4\leq m\leq 8$.
% The $SO(3)$ matrices $X^a$ take the same as the vacuum.
For simplicity,
we take an example of the vacuum 
\begin{equation}
 X^a=-\frac{\mu}{3}L_a^{[N_5]}\otimes\mathbf{1}_{N_2},
  \label{simple-vac}
\end{equation}
where $L_a^{[N_5]}$ is in the spin $(N_5-1)/2$ irreducible representation
and $N_2$ represents the multiplicity of the representation.
In this case, these two integers are related with the total matrix size 
in a simple way: $N=N_5N_2$.
Then the expectation value of a function of $M$, $\mathcal{O}(M)$, 
is calculated by
\begin{align}
 &\left\langle \mathcal{O}(M) \right\rangle_{MM}
 =\frac{1}{Z}\int
 \prod_{i=1}^{N_2}dm_{i}\,
 Z_{\rm 1-loop}(\{m_{i}\})
 \mathcal{O}(\{m_{i}\})
 e^{-2N\left(\frac{\mu}{6}\right)^3\sum_{i=1}^{N_2}N_5m_{i}^2},
 \nonumber \\
 &Z=\int
 \prod_{i=1}^{N_2}dm_{i}\,
 Z_{\rm 1-loop}(\{m_{i}\})
 e^{-2N\left(\frac{\mu}{6}\right)^3\sum_{i=1}^{N_2}N_5m_{i}^2},
 \label{matrix-int}
 \\
 &Z_{\rm 1-loop}(\{m_{i}\})=
 \prod_{J=0}^{N_5-1}
 \prod_{\substack{i,j=1\\j\neq i}}^{N_2}
 \left[
 \frac{\{(2J+2)^2+(m_{i}-m_{j})^2\} \{(2J)^2+(m_{i}-m_{j})^2\}}
 {\{(2J+1)^2+(m_{i}-m_{j})^2\}^2}
 \right]^{\frac{1}{2}},
 \nonumber
 % &\left\langle \mathcal{O}(M) \right\rangle_{MM}
 % =\frac{1}{Z}\int \prod_{s=1}^{\Lambda}
 % \prod_{i=1}^{N_2^{(s)}}dm_{si}\,
 % Z_{\rm 1-loop}(\{m_{si}\})
 % \mathcal{O}(\{m_{si}\})
 % e^{-2N\left(\frac{\mu}{6}\right)^3\sum_{s}\sum_{i}D_sm_{si}^2},
 % \label{matrix model}\\
 % &Z=\int \prod_{s=1}^{\Lambda}
 % \prod_{i=1}^{N_2^{(s)}}dm_{si}\,
 % Z_{\rm 1-loop}(\{m_{si}\})
 % e^{-2N\left(\frac{\mu}{6}\right)^3\sum_{s}\sum_{i}D_sm_{si}^2},
 % \\
 % &Z_{\rm 1-loop}(\{m_{si}\})=
 % \prod_{s,t=1}^{\Lambda}
 % %\prod_{J=|D_s-D_t|/2}^{(D_s+D_t)/2-1}
 % \prod_{J}
 % \prod_{i=1}^{N_2^{(s)}}\prod_{j=1}^{N_2^{(t)}}
 % \hspace{-5mm} {\phantom{\prod}}^{\prime}
 % \left[
 % \frac{\{(2J+2)^2+(m_{si}-m_{tj})^2\} \{(2J)^2+(m_{si}-m_{tj})^2\}}
 % {\{(2J+1)^2+(m_{si}-m_{tj})^2\}^2}
 % \right]^{\frac{1}{2}}.
 % \label{1loopdet}
\end{align}
where $m_i$ are the eigenvalues of $M$.
This expression is exact\footnote{
It is exact except for instanton effect.
However, the instanton effect is suppressed in the large-$N$ limit.
}.
The expression for a general vacuum is shown in Ref.~\cite{Asano:2012zt}.

In the following, we see how the geometries are realised
from $\phi$ in the BMN matrix model.
The brane picture in the type IIA superstring description
is a system of $N$ D0-branes on a curved background geometry,
equivalent to supergravitons on the plane-wave geometry
in the eleven-dimensional supergravity description.
However, 
it is conjectured that any vacua of the BMN matrix model
correspond to states with concentric spherical 2-branes and/or 5-branes\cite{Maldacena:2002rb};
therefore the brane picture needs to have 
D2-branes and/or NS5-branes in the IIA description. % in some way.
Thus it is natural to view the brane picture as 
a system of collective D0-branes forming D2- and/or NS5-branes.

The supergravity picture is expected to be obtained 
by the decoupling limit of these systems of branes.
In this picture,
the system is described by a classical supergravity solution,
labeled by a droplet configuration on the one-dimensional subspace.
Then the D2- and NS5-brane charges on the droplets correspond
to the number of branes in the brane picture,
and then to the $SU(2)$ representation that 
labels a vacuum in the BMN matrix model.

% the gauge/gravity duality of interest
% always has a D2-brane and/or NS5-brane in the supergravity solutions.

\subsection{The supergravity picture}
In this gauge/gravity duality,
the isometry of the supergravity solutions is 
$R\times SO(3)\times SO(6)$,
which is the same as the symmetry of the matrix model.
While part of the solutions are trivially determined by the isometry
with the topology of $\mathbb{R}\times S^2\times S^5$,
there are two non-trivial directions in ten dimensions 
in type IIA supergravity.
A general solution in the type IIA description
is written by a single function $V(r,z)$:
\begin{align}
 &ds_{10}^2
 =\left( \frac{\ddot{V}-2\dot{V}}{-V''} \right)^{1/2}
 \left\{-4 \frac{\ddot{V}}{\ddot{V}-2\dot{V}}dt^2
 -2 \frac{V''}{\dot{V}}(dr^2 +dz^2)
 +4 d\Omega_{5}^2 +2 \frac{V'' \dot{V}}{\Delta} d\Omega_2^2
 \right\},
 \nonumber \\
 &C_1
 =-\frac{(\dot{V}^2)'}{\ddot{V}-2\dot{V}} dt,
 \qquad
 C_3
 =-4\frac{\dot{V}^2 V''}{\Delta} dt \wedge d\Omega_2,
 \nonumber \\
 &B_2
 =\left( 
 \frac{(\dot{V}^2)'}{\Delta}+2z
 \right) d\Omega_2,
 \qquad
 e^{4\Phi} = \frac{4(\ddot{V}-2\dot{V})^3}{-V'' \dot{V}^2 \Delta^2},
 \label{LM solution}
\end{align}
where $\Delta = (\ddot{V}-2\dot{V})V''-(\dot{V}')^2$, and 
the dots and primes denote 
% $\frac{\partial}{\partial \log r}$ and $\frac{\partial }{\partial z}$
the partial derivatives with respect to $\log r$ and $z$, respectively.
The function $V(r,z)$ satisfies the axisymmetric Laplace equation:
$\frac{1}{r^2}\ddot V+V''=0$.

Let us take the solution corresponding to \eqref{simple-vac}
as an example.
This is interpreted as one stack of 5-branes and 2-branes.
% On the gravity side,
The function $V(r,z)$ is rewritten via
\begin{align}
 V(r,z)
 =V_0 \left( r^2z -\frac{2}{3}z^3 \right)
 +\int_{-R}^{R} dx
 \left(
 \frac{1}{\sqrt{(z-\frac{\pi N_5}{2}+ix)^2+r^2}}
 -\frac{1}{\sqrt{(z+\frac{\pi N_5}{2}+ix)^2+r^2}}
 \right) f(x),
 \label{V_and_f}
\end{align}
where a function $f(x)$ satisfies
\begin{align}
 f(x)-\frac{1}{\pi}
 \int_{-R}^{R} dx'
 %\left[
 \frac{\pi N_5}{(\pi N_5)^2+(x-x')^2}
 %\right]
 f(x')
 =-V_0N_5x^2
 +\text{const.},
 \quad
 \int_{-R}^{R} dx\, f(x)=N_2
 .
 \label{gravity-eq}
\end{align}

On the gauge-theory side (matrix-model side),
we consider the eigenvalue distribution of $\phi$.
As insertions of $\phi$ break the symmetry to $SO(2)\times SO(5)$,
the space associated with $\phi$ should be fibered 
on a point of $\mathbb{R}\times S^2\times S^5$;
hence the space corresponds to the two non-trivial directions,
$r$ and $z$ \cite{Asano:2014vba}.
In order to match both sides of the gauge/gravity duality,
we take the limit\footnote{
Note that the 't~Hooft coupling here is
$\lambda=g^2N_2=\frac{6^3}{\mu^3N}N_2\sim\frac{1}{\mu^3N_5}$ and that
the inequality $\lambda\gg N_5$ is a sufficient condition that 
typical lengths in string units are much greater than 1 \cite{Asano:2014eca}.
} where the classical supergravity approximation 
is valid \cite{Asano:2014vba,Asano:2014eca}:
$N_2\gg 1$ and $1/\mu^3 \gg N_5^2\gg 1$.
Since $\phi$ is localised at $-\tfrac{\mu}{3}L_3+\tfrac{i\mu}{6}M$
as in \eqref{localisation_vev},
we calculate the eigenvalue distribution of $M$,
which is determined by the saddle point equation in \eqref{matrix-int}.
Then, in the limit,
the saddle point equation for the eigenvalue distribution, $\rho(m)$, 
becomes
\begin{align}
 \rho (m)
 -\frac{1}{\pi}
 \int_{-m_m}^{m_m}dm'%\left[
 \frac{2N_5}{(2N_5)^2+(m-m')^2}
 %\right]
 \rho (m')
 =-\frac{2\mu^3NN_5}{6^3\pi}m^2
 +\text{const.},
 \quad
 \int_{-m_m}^{m_m} dm\, \rho(m)=N_2
 .
 \label{gauge-theory-eq}
\end{align}

Equations \eqref{gravity-eq} and \eqref{gauge-theory-eq}
are completely equivalent
under the following identification:
\begin{align}
 f(x)
 =\frac{\pi}{4}\rho(\tfrac{2}{\pi}x),
 \qquad
 % R_{S^5}^{\, 2}=
 4R
 =2\pi m_m,
 \qquad
 V_0
 =\frac{2\mu^3N}{6^3\pi^2}.
 \label{identification}
\end{align}
Therefore, we see that
the eigenvalue density of $\phi$ can reproduce 
the nontrivial part of the supergravity solution, via \eqref{identification}
through \eqref{V_and_f} and \eqref{LM solution}.
It is shown in Ref.~\cite{Asano:2014eca}
that this equivalence holds for the correspondence of any vacua 
in the matrix model, 
and hence in other $SU(2|4)$ symmetric gauge theories as well.
% part of Einstein's equations is obtained by 

\subsection{The brane picture}
In this subsection,
we consider branes in the eleven-dimensional supergravity description.
The geometry of branes in the brane picture is spherical due to the symmetry.
Its radius is determined by the light-cone classical Hamiltonian of M-branes.

Let us take an example of one M5-brane.
Then the geometry is $S^5$, and its radius \cite{Maldacena:2002rb} is
\begin{align}
 r_{\rm M5}
 =\left(
 \frac{\tilde\mu p^+}{6\pi^3T_{\rm M5}}
 \right)^\frac{1}{4}
 ,
\end{align}
% \begin{align}
%  r_{\rm M2}
%  =\frac{\tilde\mu p^+}{12\pi T_{\rm M2}}
% \end{align}
% for one M2-brane
where $p^+$ is the light-cone momentum,
$\tilde\mu$ is three-form flux
and $T_{\rm M5}$ is the M5-brane tension.

On the matrix-model side,
one obtains a vacuum corresponding to M5-branes
by taking the large-$N_2$ limit while fixing $N_5$ finite \cite{Maldacena:2002rb}.
For a single M5-brane,
the corresponding vacuum is \eqref{simple-vac} with $N_5=1$ at large $N_2$.
In the decoupling limit for an M5-brane, $N_2\gg 1$ and $1/\mu^3 \gg 1$,
the distribution of $\phi$ is dominated by $M$, and then
the solution to the saddle point equation \eqref{gauge-theory-eq}
is
\begin{align}
 \rho(m)
 =\frac{8}{3\pi m_m}\left[1-\left(\frac{m}{m_m}\right)^2\right]^\frac{3}{2},
\end{align}
with
\begin{equation}
 m_m
  % =\left(
  %   8\frac{6^3}{\mu^3N_5}
  %  \right)^\frac{1}{4}
  =\left(
    8\frac{6^3}{\mu^3}
   \right)^\frac{1}{4}
  ,
\end{equation}
for this vacuum \cite{Ling:2006up,Asano:2014vba}.
Then since $M$ corresponds to one of the $SO(6)$ matrices $X^m$
and $\phi$ should be considered as a direction
perpendicular to $\mathbb{R}\times S^2\times S^5$,
it is natural to assume
the eigenvalue distribution $\rho(m)$ is a distribution
projected onto a straight line passing through the centre of $S^5$.
Thus, we regard
the $SO(6)$ symmetric uplift of $\rho(m)$ to six dimensions
as the geometry realised by $\phi$.
% \begin{equation}
%  \int d^6\vec r\, \hat\rho(\vec r)m^n
%  =\int_{-m_m}^{m_m} dm\, \rho(m)m^n
% \end{equation}
The solution to the uplift is a spherical shell distribution
\cite{Filev:2013pza,Filev:2014jxa,Asano:2017xiy,Asano:2017nxw}:
\begin{equation}
 \hat \rho(\vec r)=
 \frac{1}{\pi^3m_m^5}\delta \left( |\vec r|-m_m \right)
 .
\end{equation}
The radius $m_m$ is measured by $X^m$ in \eqref{BMN-action}.
However, we need to rescale the matrices
to restore the original normalisation
before the matrix regularisation of the light-cone supermembranes.
By using the rescaling %of $\mu$,
% $\mu^3=(2\pi)^2l_p^6(\tilde\mu p^+)^3/N^4$,
$\tilde\mu=(\frac{N^2}{2\pi})^{\frac{2}{3}}\frac{\mu}{p^+l_p^2}$,
with $T_{\rm M5}=\frac{1}{(2\pi)^5l_p^6}$,
we obtain the rescaling of $m_m$ as
\begin{align}
 &r_0
 =\frac{2\pi\tilde\mu p^+l_p^3}{6N}m_m
 % =
 % \left(
 % \frac{\tilde\mu p^+}{6\pi^3T_{\rm M5}}
 % \right)^\frac{1}{4}
 =r_{\rm M5}
 .
\end{align}
Therefore, we confirm that, for the case of $N_5=1$,
the spherical shell distribution obtained in the matrix model
has the same radius as
the $S^5$ obtained by the light-cone Hamiltonian of an M5-brane.
For multiple M5-branes, the radius of the spherical shell distribution for a stack
is proportional to the quartic root of a light-cone momentum of one M5-brane in the stack,
instead of $p^{+\,\frac{1}{4}}$ \cite{Asano:2017nxw},
which agrees with the prediction in Ref.~\cite{Maldacena:2002rb}.

The same goes for the M2-brane.
In this case the decoupling limit is 
$N_5\gg 1$ and $1/(\mu^3N_5^2) \gg 1$
so that the distribution of $\phi$ is dominated by $-\frac{\mu}{3}L_3$,
at least at large $N_2$\footnote{
In the M2 case,
$N_2$ needs to be large unlike the M5 case
so that the instanton effect is suppressed.
However, if the instantons do not affect 
the dominance of $-\frac{\mu}{3}L_3$ in $\phi$,
the statement that $\phi$ reproduces the predicted spherical shell distribution
is valid even at finite $N_2$.
}.
Then the $SO(3)$ symmetric uplift of $\phi$ correctly reproduces 
the spherical shell distribution with the correct radius.

\section{Thermal BMN Matrix Model}
We now have a fairly good picture of emergent geometries.
However, it is far non-trivial 
how these geometries emerge when we scale energy or temperature.
In this section, we consider the BMN matrix model with temperature
and briefly review 
its gravity dual. % for the BMN matrix model 
% changes by the temperature.
Finally we see qualitative agreement with lattice simulation.

\subsection{Gravity side}
While the supergravity solutions without temperature
is known as the Lin-Maldacena geometry \eqref{LM solution},
discussed in the previous section,
its thermal version has not been completely obtained yet.
Fortunately, 
a solution in which the black hole horizon has the simplest topology\footnote{
$S^1$ here is the M-theory circle},
$S^1\times S^8$,
% dual to the deconfined phase of the BMN matrix model at strong coupling
was constructed with some numerical computations
%perturbatively in terms of derivatives
\cite{Costa:2014wya}.

As we are interested in the gauge/gravity duality,
the solution is for the strong coupling regime, i.e.~$\mu\to 0$.
Then a finite parameter of the solution is $\mu/T$.
At small $\mu/T$,
the solution is approximated by the non-extremal black 0-brane,
which has $R\times SO(9)$ symmetry.
Then the solution in Ref.~\cite{Costa:2014wya} is numerically obtained
by a continuous deformation by $\mu/T$,
which breaks the symmetry and 
makes the solution asymptotically approach
the plane-wave geometry with $R\times SO(3)\times SO(6)$ symmetry 
at infinity.
Thus the black hole horizon of the solution has the topology of $S^8$.
%$S^1_M\times S^8$, where $S^1_M$ is the M-theory circle.
On the gauge-theory side, this corresponds to the deconfinement phase 
with thermal fluctuations around the trivial vacuum, $X^a=0$.

The free energy is 
\begin{align}
 F(T,\mu)
 =-c_1N^2T^{\frac{14}{5}}
 %F_{\rm BFSS}(T)
 f(\tfrac{\mu}{T}),
\end{align}
where $c_1=\frac{1}{21}(\frac{120\pi^2}{49})^{\frac{7}{5}}$.
The factor $-c_1T^{\frac{14}{5}}$ is the free energy of 
the non-extremal black 0-brane, 
which corresponds to the BFSS matrix model
\cite{Klebanov:1996un,Itzhaki:1998dd}.
The function $f(\mu/T)$ is the one numerically obtained,
which turns out to be a monotonically decreasing function.
Since, in the matrix model,
the free energy behaves as $N^2$ in the deconfinement phase
and as $N^0$ in the confinement phase,
the corresponding transition on the gravity side occurs at $f(\mu/T)=0$
in a way similar to the Hawking-Page transition \cite{Hawking:1982dh}.
The solution to $f(\mu/T)=0$ is 
\begin{equation}
 T_c/\mu=0.105905(57),
  \label{gravity-prediction}
\end{equation}
and this gives an upper bound of the critical temperature.

\subsection{Lattice simulation}
In this subsection, we look at two types transitions
in the thermal BMN matrix model,
observed in the rational hybrid Monte Carlo simulation 
in Ref.~\cite{Asano:2018nol}.
The number of lattice sites in the results shown in this paper
is $\Lambda=24$.
We measure the following observables:
\begin{align}
 &\left\langle \vert P\vert \right\rangle
 =\left\langle \frac{1}{N} \Big\vert
 \Tr \left( \exp \left[ i \beta A \right] \right)
 \Big\vert \right\rangle \, ,
 \label{PolyakovLoop}\\
 &{\rm Myers}
 =\left\langle \frac{i}{3N\beta}\int_0^\beta d\tau \epsilon_{abc}\Tr(X^aX^bX^c) \right\rangle \, ,
 \label{Myers}\\
 &R^2_{ii} 
 =\left\langle \frac{1}{N\beta}\int_0^\beta d\tau \Tr(X^{i}X^{i})\right\rangle
 \qquad \text{(no sum on $i$)},
 \label{R2ii}
\end{align}
which are the Polyakov loop, the Myers term and the extent of the eigenvalue distribution of $X^i$ in each direction, respectively.
Here, $\beta$ is the inverse temperature,
$A$ in \eqref{PolyakovLoop} is the gauge field in the uniform gauge,
and $i$ in \eqref{R2ii} runs from 1 to 9.

\begin{figure}[t]
 \centering
 \begin{tabular}{@{}c@{}}
 \includegraphics[scale=0.55]{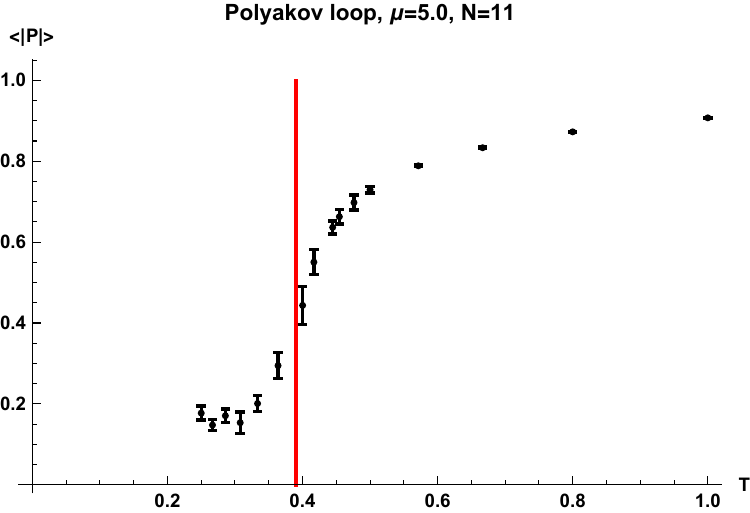}
 \includegraphics[scale=0.55]{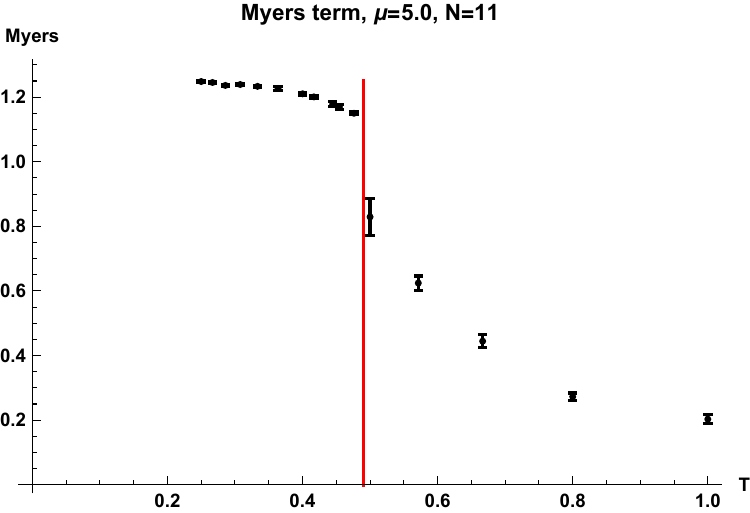}
 \end{tabular}
 \caption{\small The Polyakov loop (top) and the Myers term (bottom) for $\mu=5.0$, $N=11$.
 The transition temperatures of the Myers term and of the Polyakov loop are around $T_{c1}=0.490$ and $T_{c2}=0.390$, respectively.
 }
 \label{fig:N11mu5L24}
\end{figure}

\begin{figure}[t]
 \centering
 \begin{tabular}{@{}c@{}}
  \includegraphics[width=2.74in]{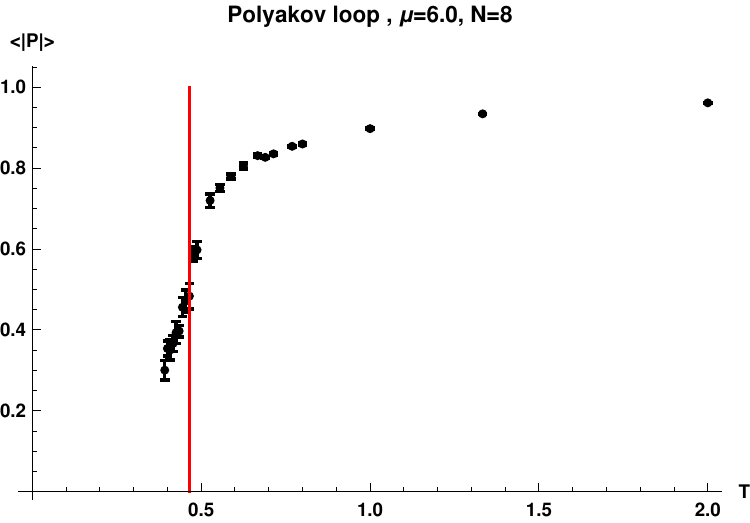}
  \includegraphics[width=2.74in]{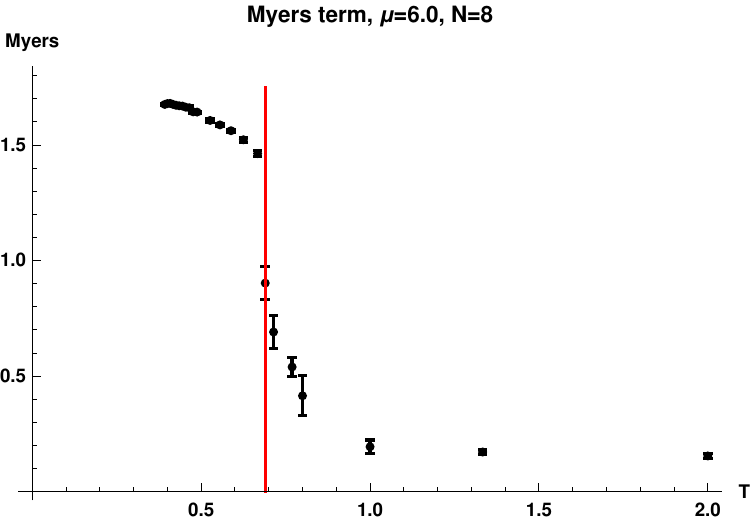}
 \end{tabular}
 \includegraphics[width=2.74in]{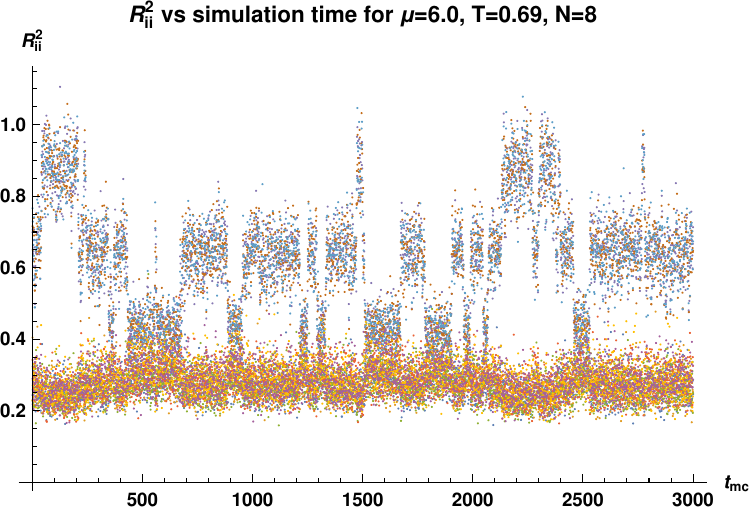}
 \caption{\small The Polyakov loop (top left), the Myers term (top right) and $R^2_{ii}$ at $T_{c1}=0.690$ (bottom) for $\mu=6.0$, $N=8$.
 The transition temperatures of the Myers term and of the Polyakov loop are around $T_{c1}=0.690$ and $T_{c2}=0.465$, respectively.
 The upper bluish points in the plot for $R^2_{ii}$ correspond to the $SO(3)$ matrices and the lower reddish points correspond to the $SO(6)$ matrices.}
 \label{fig:N8mu6L24}
\end{figure}

Fig.~\ref{fig:N11mu5L24} 
and the top graphs in Fig.~\ref{fig:N8mu6L24} respectively
show two transitions observed with $N=11$ and $8$.

The left graphs in Fig.~\ref{fig:N11mu5L24} and \ref{fig:N8mu6L24}
show a transition in the Polyakov loop.
% which we call the Polyakov-loop transition in this paper.
Since the Polyakov loop is an order parameter for 
the confinement/deconfinement transition,
this transition involves a change of the eigenvalue distribution of 
the gauge field from the uniform one to a non-uniform one.
At this point, it is not clear yet 
whether there exists only one transition or more transitions 
in the transition region in the Polyakov loop;
however, at least a gapped distribution is observed at high temperatures
and an ungapped one at temperatures slightly lower than 
the transition temperature marked as the red vertical line 
for the Polyakov loop.
Note that the transition observed in the numerical simulation is merely effective 
because the true phase transition exists only in the large-$N$ limit.

On the other hand, 
the right graphs in Fig.~\ref{fig:N11mu5L24} and \ref{fig:N8mu6L24} 
show another transition,
detected by the Myers term\footnote{
The Myers transition can be equally detected, for example, by 
the extent of the $SO(3)$ $X^a$: $\sum_{a=1}^3R^2_{aa}$.
},
which we call the Myers transition in the paper.
One can see the two transition temperatures are distinct
by comparing the positions of two red vertical lines
in the right and left graphs.
In fact, what the Myers term captures is completely different.
The bottom graph in Fig.~\ref{fig:N8mu6L24} is 
a Monte-Carlo history of $R^2_{ii}$ 
at the transition temperature $T_{c1}=0.690$ 
marked as the red vertical line in the top right graph.
It shows large fluctuations between three different levels 
for the $SO(3)$ matrices, $X^a$.
Each level corresponds to one or more vacua of the BMN matrix model;
hence the fluctuations in the plot are transitions between vacua.
Since these levels are what the Myers term measures,
the transition in the Polyakov loop 
is essentially different from the Myers transition.

As the temperature decreases, the system undergoes the Myers transition
before the confinement/deconfinement transition.
At higher temperatures, the system is in %a ``matrix phase,''
a phase where thermal fluctuations are around the trivial vacuum.
With decreasing temperature, 
the system enters the Myers transition region
and then hops among different vacua, seen in Fig.~\ref{fig:N8mu6L24}.
After the transition,
it stays at a certain vacuum with non-zero $SU(2)$ spins.
Let us call it a fuzzy-sphere phase.
Spins of a vacuum in the final phase vary by value of $\mu$;
the smaller the value of $\mu$, the larger a typical value of spins.

\begin{figure}[htbp]
 \centering
 \begin{tabular}{@{}c@{}}
  \includegraphics[width=2.74in]{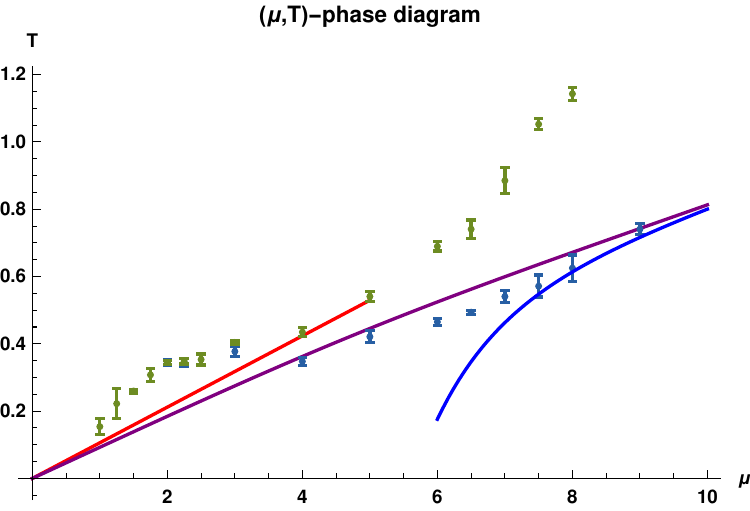}
  \includegraphics[width=2.74in]{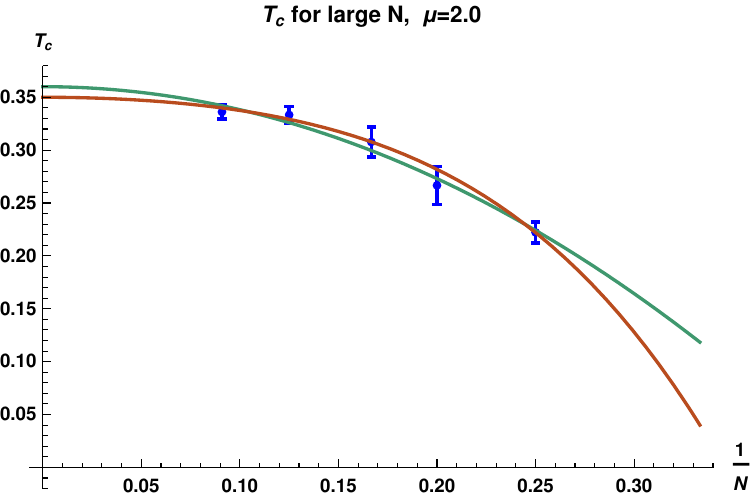}
 \end{tabular}
 \caption{\small The observed phase diagram in the $\mu$--$T$ plane for $N=8$ and $\Lambda=24$ (left) and large-$N$ extrapolation of the transition temperature of the Myers transition at $\mu=2.0$ (right).
 On the left figure, the green and dark blue points respectively represent the Myers transition and the Polyakov-loop transition.
 The blue curve is the large-$\mu$ expansion of the critical temperature to three-loop order,
 the red line is the gravity prediction,
 and the purple curve is the Pad\'e approximant obtained from the large-$\mu$ expansion.
 On the right figure, observed data are of $N=4,5,6,8$ and $11$.
 The green curve is a quadratic extrapolation: $0.36-2.17/N^2$
 while the red curve is a quartic one: $0.35-1.09/N^2-15.26/N^4$.
 % which gives $T_c(\infty)=0.35\pm0.01$.
 }
 \label{fig:phase-diagram}
\end{figure}

The left graph in Fig.~\ref{fig:phase-diagram}
is the resultant phase diagram for $N=8$ with $\Lambda=24$.
Though not presented here,
we confirmed the perfect agreement with the large-$\mu$ expansion of the critical temperature \cite{Spradlin:2004sx,Hadizadeh:2004bf},
at large $\mu$ such as $\mu=14,18$.
Surprisingly, the two transitions appear to merge around $\mu=3$.
Although it is difficult to determine the transition temperature 
in the Polyakov loop at $\mu\lesssim 2$,
we can estimate it approximately, and 
it seems the two transition temperatures coincide.
Thus we naturally assume 
the critical temperature of the confinement/deconfinement transition 
is equivalent to that of the Myers transition in this region.
Then the transition temperature seems to approach the theoretical prediction from the gravity side.

Although the matrix size is fixed to $N=8$ in the phase diagram,
it is seen 
that the finite-$N$ effect is negligible at $\mu=2.0$,
from the large-$N$ extrapolation of the transition temperature,
shown in the right graph of Fig.~\ref{fig:phase-diagram}.

\section{Summary and Discussion}
As to the emergent geometries embedded in the BMN matrix model,
we have seen that both
the geometries in the supergravity picture and 
those in the brane picture are reproduced 
by the eigenvalue distribution of the BPS operator $\phi$.

In the supergravity picture,
% The infinitely many discrete vacua in the large-$N$ BMN matrix model 
% correspond to supergravity solutions characterised by
% droplets of smeared M2 and M5 charges.
the supergravity solutions are described by a single function $V(r,z)$.
Through the gauge/gravity dictionary \eqref{identification},
this function can be constructed by 
the eigenvalue density of $\phi$ on the gauge-theory side.

In the brane picture,
the geometry of M2-branes or M5-branes is a two-sphere or a five-sphere,
respectively.
Its radius is determined by the light-cone Hamiltonian of M-branes.
Then, by the symmetric uplift of the eigenvalue distribution of $\phi$,
a sphere wrapped by M2-branes or M5-branes is reproduced
with the expected radius.
% Note that the realisation of $S^5$ of M5-branes and $S^2$ of M2-branes by the eigenvalues of $\phi$ are in a consistent way.
For M2-branes,
this is understood as the Myers effect \cite{Myers:1999ps} as well.

It provides a unified view that 
a single BPS operator $\phi$ describes both 
the supergravity solution and the brane geometry.
However,
it still remains obscure how exactly this equivalence 
of the matrix model to the brane picture
is related with the equivalence to 
the supergravity picture in the gauge/gravity duality.
If the equivalence among these three pictures is established
in the BMN-model case,
one can expect deeper understanding of the gauge/gravity conjecture in general.

Another important aspect is relationship to the non-commutative geometry.
In the case of the 2-brane,
the emergent $S^2$ geometry %in the brane picture 
is constructed by a fuzzy 2-sphere %made from the $su(2)$ algebra
as a result of the Myers effect.
Meanwhile, for the 5-brane,
a concrete construction of $S^5$ 
by a set of matrices in the BMN model as a non-commutative sphere 
has not been obtained yet.
The difficulty stems from the fact that
the $S^5$ is realised thanks to the quantum effect 
while the realisation of the $S^2$ can be seen even at the classical level.
Although the BPS operator $\phi$ overcomes the difficulty of the quantum effect
thanks to the supersymmetry,
it fails to offer a matrix form of the geometry at this point.

The numerical results showed an intriguing insight
into a dynamical aspect of the emergent geometries, namely, 
a geometrical interpretation given by the Myers transition.
Since each vacuum corresponds to a supergravity solution,
the Myers transition represents a transition between supergravity solutions.
At high temperatures, the system is around the trivial vacuum, 
the spin-0 representation.
% which corresponds to the supergravity solution of a single 5-brane.
Since this is in the deconfinement phase,
the corresponding gravity solution is the one in Ref.~\cite{Costa:2014wya}.
As the temperature decreases, the system hits the phase boundary of 
the Myers transition
so that the system is transitioned to a fuzzy-sphere vacuum,
which corresponds to another supergravity solution.
Until it undergoes another phase transition---the confinement/deconfinement 
transition---the gravity solution corresponding to the phase would be 
a Lin-Maldacena-like black-hole geometry with a non-trivial horizon topology,
which has not been discovered yet.
Then as the temperature decreases further,
the corresponding gravity solution should become a thermal Lin-Maldacena geometry
without a horizon.
However, it should require more to understand the exact meaning of these interpretations.
In order for the classical supergravity picture or the brane picture to be valid,
one needs to take a proper large-$N$ limit;
$N\approx 10$ may be insufficient to see geometries.
In addition,
it can be obscure whether the high-temperature phase has a good geometrical meaning
at intermediate $\mu$
because at high temperatures it is described by the perturbative matrix model
rather than by the gravity solution\footnote{
As the gravity picture is valid at small $\mu$ with $T/\mu$ fixed,
% a set of parameters of the thermal BMN model is $\mu$ and $T$ in this paper,
large $T$ means large $\mu$, which is a weak coupling.
}.
% the thermal effect would mask geometrical features;
% in this sense the Myers transition would look like ``no geometry $\to$ geometry.''%

Another remark on the emergent geometries observed in the numerical simulation
is the Myers transition at small $\mu$ looks like a transition to 
a vacuum corresponding to 5-branes.
The observation was the Myers term in the fuzzy-sphere phase looked 
independent of $N$ \cite{Asano:2018nol}.
In the simple setup \eqref{simple-vac},
the Myers term \eqref{Myers} behaves as
$\sim \mu^3(N_5^2-1)$.
% $\mathrm{Myers}\sim \frac{\mu^3}{3^4N}\Tr(L^aL^a)\sim \mu^3(N_5^2-1)$.
Thus it is likely that a typical dimension of irreducible representations does not grow as $N$ increases
while a typical multiplicity grows linearly,
which suggests the phase of such a vacuum
corresponds to 5-branes \cite{Maldacena:2002rb}.

The two observed phase transitions
give different transition temperatures in the intermediate region of $\mu$.
The numerical simulation revealed that
these two transitions do not merge at $3 \lesssim \mu \lesssim 6$
for $N=8$ on the lattice
while it appears they merge below $\mu\approx 3$.
There is no theoretical explanation of this behaviour thus far.

We conclude the simulation result is 
consistent with the gravity prediction obtained in Ref.~\cite{Costa:2014wya}.
The numerical result in Fig.~\ref{fig:phase-diagram} shows qualitative agreement with 
the expected critical temperature of the confinement/deconfinement transition 
\eqref{gravity-prediction}.
% and the numerical result of the matrix model.
Strictly speaking,
we need to keep the system around the trivial vacuum 
below the transition temperature of the Myers transition
for more rigorous agreement
because the obtained gravity prediction corresponds to the trivial vacuum.
Since all vacua are stable at zero temperature,
this seems possible in practice
if one manipulates a simulation so as not to let the system undergo the Myers transition.
However, as the confinement/deconfinement transition 
seems to merge with the Myers transition at small $\mu$,
the obtained transition temperature should be a reasonable
estimate of the critical temperature for the trivial vacuum.
% that the transition temperature in the Polyakov loop
% is dependent on vacua the system is around,
Moreover, it was found
that the trivial vacuum gives a higher transition temperature 
in the Polyakov loop than fuzzy-sphere vacua \cite{Asano:2018nol}.
This observation is also consistent with the prediction
that \eqref{gravity-prediction} is an upper bound.

There are much more to study numerically in the low-temperature region.
Since the gravity side at zero temperature is described by droplet solutions,
we expect a richer structure at lower temperatures,
which should reflect geometrical information.
It would be a real challenge to reveal the complete phase structure 
due to more computational cost.
This is because, at lower temperatures, the lattice effect becomes larger
and, more importantly, one needs to carefully deal with 
contributions from all vacua
as they are expected to be (meta-)stable.
Further numerical studies in this region will enhance
our understanding of dynamics of emergent geometries.

There are other intriguing directions related with the BMN-model simulation.

One example is the bosonic BMN matrix model, 
the action of which has only the bosonic part of the full BMN matrix model.
In Ref.~\cite{Asano:2020yry},
we found that there are two transitions at finite $N$ in the eigenvalue distribution of the gauge field---a uniform-to-non-uniform transition
% a transition from the uniform distribution to an ungapped one
% and a transition from an ungapped distribution to a gapped one
and an ungapped-to-gapped transition---but also
that these two converge to a single first-order transition in the large-$N$ limit.
This may suggest the transition observed in the Polyakov loop in the full BMN matrix model
is also a single first-order transition.
The behaviour of the Polyakov loop
is further studied in terms of counting states 
in an upcoming paper (see Ref.~\cite{Kovacik:2020cod}).

Another direction is to study matrix models for longitudinal M5-branes.
While M5-branes appearing in the BMN matrix model are transverse to the
M-theory direction, 
those in the Berkooz-Douglas (BD) matrix model \cite{Berkooz:1996is},
which is the BFSS matrix model probing D4-branes,
are longitudinal to the direction.
There have been numerical results of the BD model,
and they showed the agreement in the gauge/gravity duality
\cite{Filev:2015cmz,Asano:2016kxo}.
More interestingly, there is proposed a BMN-like mass-deformed version of 
the BD matrix model \cite{Kim:2002cr},
which is supposed to describe 
M-theory on the plane-wave background probing longitudinal M5-branes.

\acknowledgments
% ``Workshop on Quantum Geometry, Field Theory and Gravity''
The author thanks the organisers of the Corfu Summer Institute 2019 for its hospitality
and Veselin Filev, Goro Ishiki, Samuel Kov\'a\v{c}ik, Denjoe O'Connor, 
Takashi Okada, Shinji Shimasaki and Seiji Terashima
for the collaborations presented in this paper.
The numerical results were obtained on Fionn and Kay
at the Irish Centre for High-End Computing (ICHEC).
The author is supported by the JSPS Research Fellowship for Young Scientists.

% \bibliographystyle{JHEP}
% \bibliography{201909CorfuRef}

\end{document}